\documentclass[preprint]{template/vgtc}

\graphicspath{{figures/}{pictures/}{images/}{./}}

\usepackage{times}

\usepackage{mathptmx}
\usepackage{placeins}
\usepackage{float}
\usepackage{tikz}
\usetikzlibrary{arrows.meta,positioning}

\definecolor{paperblue}{HTML}{8FB8C9}
\definecolor{paperbluepale}{HTML}{D9EBF1}
\definecolor{paperamber}{HTML}{F9D991}
\definecolor{paperpurple}{HTML}{E8DDF0}
\definecolor{papergray}{HTML}{E2E6E8}

\preprinttext{}

\vgtcinsertpkg

\title{Beyond Activation: Gaze Invocation with Visible Status for an Embodied AR Assistant in Co-Located Collaboration}

\author{Chenrui Ma\thanks{e-mail: machenrui233@gmail.com}\\
        \scriptsize Graduate School of Interdisciplinary Information Studies\\
        \scriptsize The University of Tokyo, Tokyo, Japan
\and Qing Zhang\thanks{Corresponding author. e-mail: qzkiyoshi@gmail.com}\\
        \scriptsize Interfaculty Initiative in Information Studies\\
        \scriptsize The University of Tokyo, Tokyo, Japan
\and Yoshio Ishiguro\thanks{e-mail: ishi@acm.org}\\
        \scriptsize Interfaculty Initiative in Information Studies\\
        \scriptsize The University of Tokyo, Tokyo, Japan}

\hypersetup{
  pdftitle={Beyond Activation: Gaze Invocation with Visible Status for an Embodied AR Assistant in Co-Located Collaboration},
  pdfauthor={Chenrui Ma, Qing Zhang, Yoshio Ishiguro},
  pdfsubject={}
}

\teaser{
  \centering
  \begin{minipage}[c]{.34\linewidth}
    \centering
    \includegraphics[width=\linewidth]{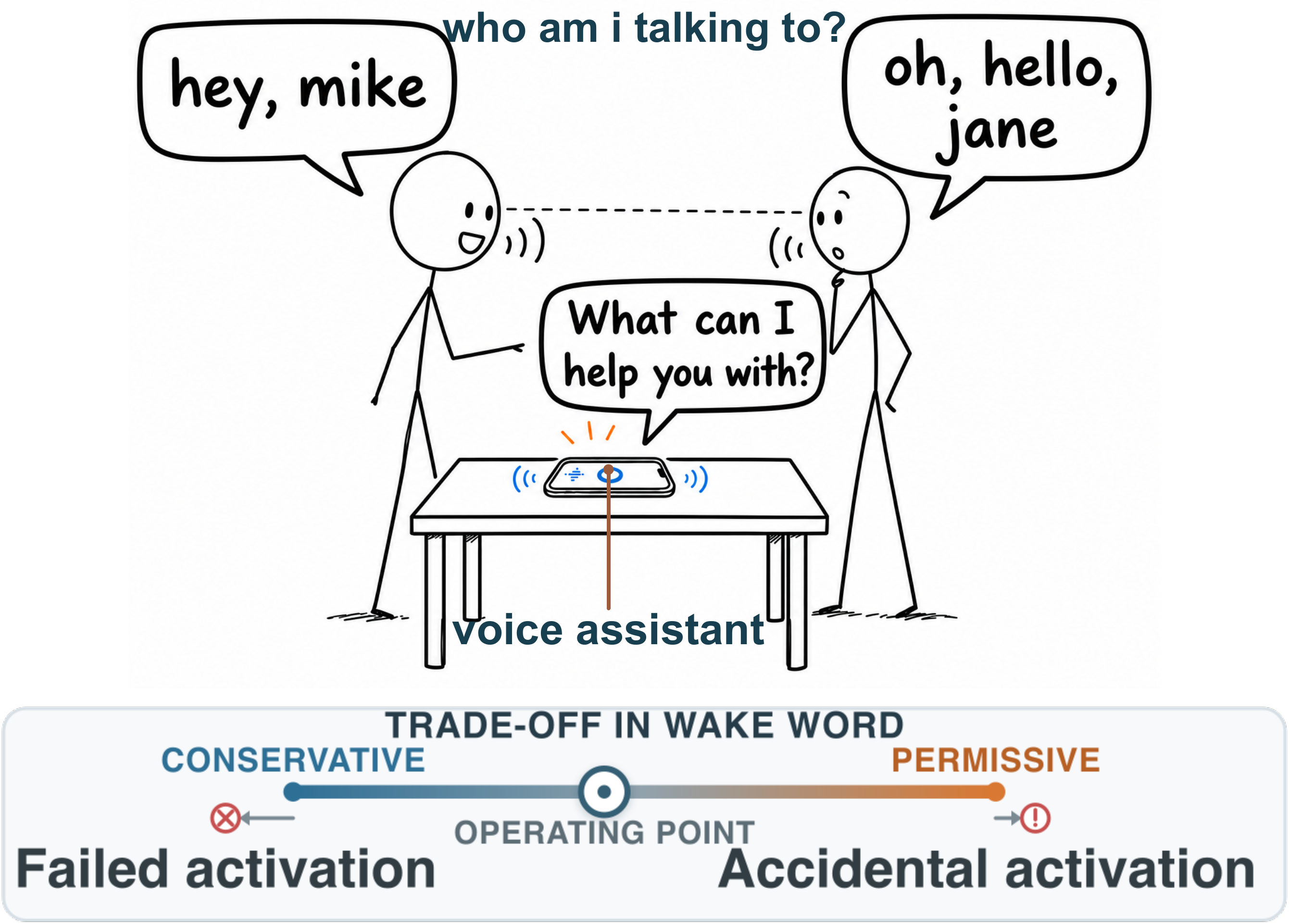}
  \end{minipage}\hfill
  \begin{minipage}[c]{.21\linewidth}
    \centering
    \includegraphics[width=\linewidth]{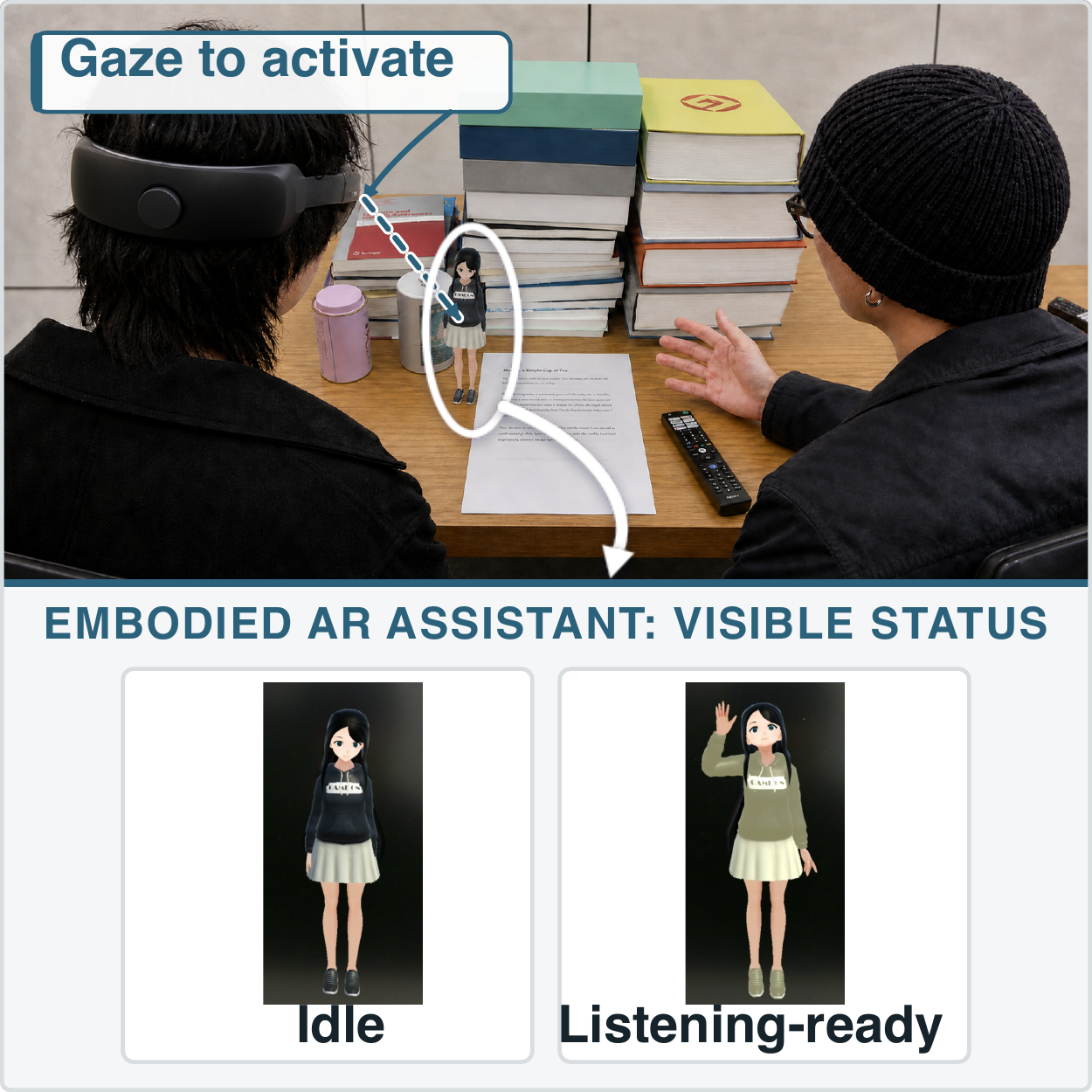}
  \end{minipage}\hfill
  \begin{minipage}[c]{.42\linewidth}
    \centering
    \includegraphics[width=\linewidth]{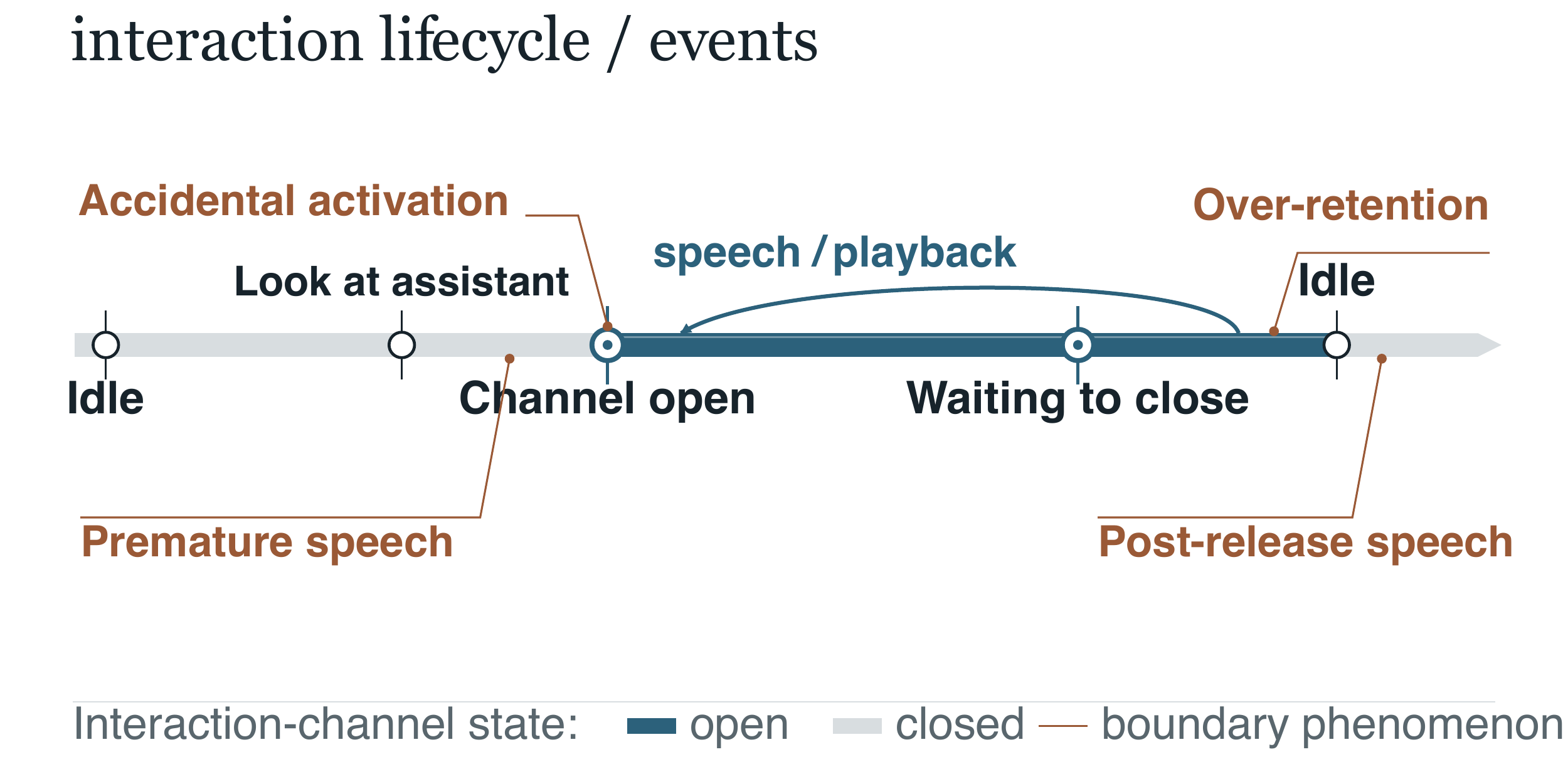}
  \end{minipage}
  \caption{From wake-word ambiguity to visible gaze invocation. Left: increasing
  wake-word sensitivity reduces missed commands but also admits more accidental
  triggers. Middle: the wearer addresses a spatially anchored AR assistant whose
  status is visible. Right: gaze opens the interaction channel, speech and
  playback keep it open, and inactivity closes it.}
  \label{fig:teaser}
}

\abstract{
Wake words face an inherent trade-off: higher sensitivity reduces missed
commands but increases accidental activations. In co-located augmented reality
(AR), the system must also determine whether the user is speaking to the
assistant or to a nearby person. We introduce a new perspective on this
problem: combining gaze-based address with an embodied assistant, visible
listening status, and turn management across activation, continued
interaction, and release. In our implementation, users look at an assistant
anchored in the scene and see activation progress and listening status on its body.
Sustained gaze opens a local interaction channel; speech and playback keep it
open when attention returns to the task; and inactivity closes it. We evaluated this design with 25
participants. After selecting the assistant's placement and dwell duration,
each participant worked with a partner to plan a trip while using the
assistant. The study recorded 500 interaction outcomes, including 496
assistant-directed utterance attempts. Gaze acquisition completed before speech
for 490 of these attempts (98.8\%): 467 began while the channel remained open,
whereas 23 began after it had been released. The other 6 attempts began before
acquisition completed. Among 102 reviewed attempts in which gaze left after
acquisition but before speech, the configured retention policy kept the channel
open for 79 and released it before 23.
The results show that an embodied target with visible
status can support clear entry into an assistant interaction while revealing a
different problem at release: after users look back to their work, they may not
see that the assistant has stopped listening. We contribute the implemented
gaze-invocation design and design implications for communicating assistant
state after visual attention moves elsewhere.
}


\begin{document}

\firstsection{Introduction}

\maketitle

Intelligent assistants need a reliable way to tell whether nearby speech is
meant for them, which is a recurring source of breakdown in everyday use
\cite{porcheron2018voice,beneteau2019communication,luger2016like}. Wake words
provide an explicit signal, but no single sensitivity setting avoids both
failure modes: more permissive detection catches more intended commands and
also admits more background or conversational speech; stricter detection
reduces accidental triggers and misses more intended commands
\cite{michaely2017keyword}. When users work with an AR assistant and another
person at the same time, their speech may be directed to either one while they
handle physical materials \cite{auer2018gaze,bohus-horvitz-2011-multiparty}.

Existing systems address parts of this ambiguity through spatial and
multimodal cues. Tama uses gaze toward an embodied smart-speaker head, while
Look and Talk and Look to Wake use sustained gaze toward AR assistants
\cite{mcmillan2019Tama,zhang2025look,shiraishi2025look}. Persistent Assistant
combines gaze and pinch to connect repeated queries
\cite{cho2025persistent}. Other approaches infer whether an HMD wearer is
addressing another person or combine gaze, pointing, and dialogue history to
resolve spoken references after a trigger phrase
\cite{yi2024interactive,lee2024gazepointar}. These systems show that gaze,
embodiment, and multimodal context can clarify activation, addressee selection,
or reference. However, gaze-invocation studies primarily examine whether and
how an interaction starts, while addressee and reference systems solve adjacent
classification problems. They do not jointly examine how a visibly embodied
assistant should remain available after activation, decide when the interaction
has ended, and communicate that transition while the wearer is collaborating
with another person.

This gap matters because activation and release occur under different visual
conditions. An embodied assistant anchored in the physical scene gives users a
stable target to look at and a place where the system can display its status.
During activation, the user is already looking at that display. Afterward,
people often look back to documents or toward a collaborator while planning
what to say
\cite{auer2018gaze,glenberg1998averting,andrist2014conversational,
bogels2015neural}. Requiring continuous gaze would interrupt this behavior, but
once gaze moves elsewhere, status displayed only on the assistant may no
longer be seen. A gaze-invocation method therefore needs to manage not only
entry but also continued availability and communication of release.

We address this need by separating three stages of the interaction: sustained
gaze opens the channel (\emph{acquisition}); speech and assistant playback keep
it open while gaze moves elsewhere (\emph{retention}); and a period without
relevant activity closes it (\emph{release}). Because looking alone is
ambiguous---the classical Midas-touch problem---we use dwell to filter shorter
incidental looks
\cite{jacob1990what,kumar2007eyepoint,mohan2018dualgaze}. A short dwell responds
quickly but is easier to trigger accidentally; a long dwell reduces incidental
activation but delays access
\cite{majaranta2009fast,fernandez2020gazewheels,isomoto2022interaction}. Our
design therefore lets each participant select the dwell duration and displays
activation progress and readiness on the assistant. Opening status is easy to
see because the user is looking at the assistant; closing status may be missed
because release usually occurs after attention has moved away.
Figure~\ref{fig:teaser} summarizes this motivation and the
acquisition--retention--release lifecycle.

We implemented this design on HoloLens~2 and evaluated it with 25 participants.
The first three tasks established assistant placement, measured gaze-to-speech
timing, and let participants choose their dwell duration. Participants then
used the configured assistant during a co-located travel-planning task with a
collaborator. Of 496 assistant-directed utterance attempts, only 6 were
\emph{Premature speech} events in which users spoke before gaze acquisition
completed and before the ready indicator appeared. A further 23 were
post-release events: acquisition had completed earlier, but users spoke after
the assistant had returned to Idle. First-person video, a visible gaze cursor, and
assistant status cues made these transitions observable during critical
incident analysis \cite{flanagan1954critical}.

The paper makes three contributions:
\begin{itemize}
    \item \textbf{An embodied gaze-invocation design} in which the assistant is
    both the gaze target and the status display. Participant-selected dwell
    opens the interaction channel, speech and playback retain it, and inactivity
    releases it (Sections~3 and~4).
    \item \textbf{An implemented HoloLens~2 system} that keeps interaction
    control on the headset while maintaining a persistent connection to the
    conversational service, with participant-specific placement and dwell
    calibration (Sections~4 and~5).
    \item \textbf{New empirical findings throughout the invocation lifecycle.}
    Our collaborative evaluation reveals that gaze behavior and status
    visibility differ between acquisition and release, a pattern that
    activation-only evaluations overlook. These findings provide lifecycle-level
    design insights for gaze-invocation systems (Sections~6 and~7).
\end{itemize}

\section{Related Work}

\subsection{Wake-Word Invocation and Other Address Cues}

Always-on voice interfaces use keyword spotting to decide when to process
subsequent speech. Increasing sensitivity reduces missed commands but accepts
more background speech; stricter detection reduces accidental triggers but
misses more intended commands \cite{michaely2017keyword}. In shared
environments, the system must also decide whether speech addresses the device
or another person \cite{beneteau2019communication,luger2016like}. Wake-word
invocation therefore combines speech detection with addressee selection.

Gaze provides a spatial cue for addressing an assistant. Tama combines gaze
with an embodied smart-speaker head and compares it with a conventional
speech-agent interface \cite{mcmillan2019Tama}. Its video analysis identified
failures before users established gaze and failures caused by looking away;
the wake-word condition produced more successful interactions, while
mutual-gaze feedback improved the gaze condition. Look and Talk uses sustained
gaze toward an AR avatar and includes a small exploratory wake-word comparison
\cite{zhang2025look}. It reports early findings on naturalness,
unobtrusiveness, dwell duration, and activation feedback. Look to Wake uses
sustained gaze to invoke an assistant without a spoken wake word
\cite{shiraishi2025look}. These studies primarily examine whether gaze starts
an interaction and how users experience that trigger. Our work follows the
interaction after activation, including when the channel stays open, closes,
and must be opened again while the user looks elsewhere.

Related systems study other parts of assistant interaction in shared settings.
Speak-to-shoulder detects which person a head-mounted display (HMD) wearer is
addressing and reports detector accuracy and false triggers on real data
\cite{yi2024interactive}. GazePointAR combines gaze, pointing, and conversation
history to resolve spoken references, while a trigger phrase still starts the
assistant interaction \cite{lee2024gazepointar}. Persistent Assistant uses gaze
and pinch to connect repeated queries and combines haptic and speech feedback
\cite{cho2025persistent}. These systems show how spatial cues, context, and
multimodal feedback can clarify an addressee or referent. Our study begins with
gaze-based selection of the embodied assistant and follows how visible status
and turn management support the interaction through retention and release.

The present design brings together gaze toward a visible assistant, status
displayed on that assistant, and control of an open interaction channel.
Acquisition, retention, and release describe how this specific gaze interaction
starts, continues, and ends. The collaborative evaluation follows all three
stages while users move their attention between the assistant, task materials,
and a collaborator.

\subsection{Gaze and Embodiment as Spatial Address}

GazePointAR illustrates the referential use of gaze in wearable AR: it combines
gaze, pointing, and conversation history to resolve spoken pronouns
\cite{lee2024gazepointar}. This determines \emph{what} the user is referring
to. Assistant invocation must also determine \emph{who} the user is addressing.

Natural looking and intentional gaze input use the same signal, creating the
Midas-touch problem \cite{jacob1990what,kumar2007eyepoint,
mohan2018dualgaze}. Common solutions include dwell or fixation, gaze gestures
or pursuit, and multimodal confirmation
\cite{majaranta2009fast,isomoto2022interaction,
fernandez2020gazewheels,namnakani2023comparing,pfeuffer2016gaze}.
Shorter dwell speeds activation and admits more incidental looks; longer dwell
reduces accidental activation but delays intended access
\cite{majaranta2009fast,fernandez2020gazewheels}.
Adaptive dwell can vary the threshold with inferred intent, attention, or task
context, but a changing trigger time can weaken the user's sense of control
\cite{narkar2024gazeintent}. GazeIntent users still preferred static dwell in
some tasks \cite{narkar2024gazeintent}, while a focus-adaptive EyeBCI condition
produced more fatigue and frustration than fixed-dwell conditions
\cite{barresi2016focus}. We
therefore let each participant select a dwell duration and hold it constant
during the collaborative evaluation.

An embodied AR assistant provides a stable gaze target and a place to display
feedback. Prior work shows that agent form affects perception and task fit
\cite{cassell2001embodied,kim2018body,wang2019exploring,
reinhardt2020embedding,hajahmadi2025voice,schlager2024nonhumanoid}, and
nonverbal cues communicate attention and coordination
\cite{urakami2023nonverbal,andrist2017looking}. Status displayed on the agent is
easy to see during gaze activation because the user is already looking at it.
Peripheral, audio, and haptic cues provide additional ways to communicate
changes after attention moves elsewhere \cite{lee2024may,pfeuffer2016gaze}.

\subsection{Conversational Coordination and Control of the Assistant's Turn}

Conversation and human--robot interaction (HRI) research use gaze for addressee
selection, floor management, and turn allocation \cite{auer2018gaze,
andrist2014conversational,admoni2017social}. Situated dialogue systems similarly
combine gaze, speech, timing, feedback, and joint attention to manage
participation \cite{bohus-horvitz-2011-multiparty,
almoubayed2011turntaking,skantze2021turntaking,skantze2014jointattention,
andrist2017looking}. Embodied agents can mediate collaborative decisions
\cite{han2026mediation}, while social virtual-reality (VR) systems use visual
and spatial-audio cues to direct attention toward a potential human speaker
\cite{lee2024may}. This research shows that turn-taking depends on gaze, speech,
feedback, and timing. Our design applies these principles to gaze activation
for an embodied AR assistant: speech and playback keep an open channel
available, inactivity closes it, and the assistant displays the controller's
current state.

\section{Gaze Invocation with Visible Assistant Status}

\begin{figure*}[!t]
    \centering
    \begin{tikzpicture}
      \node[anchor=south west,inner sep=0] (architecture) at (0,0)
        {\includegraphics[width=\textwidth]{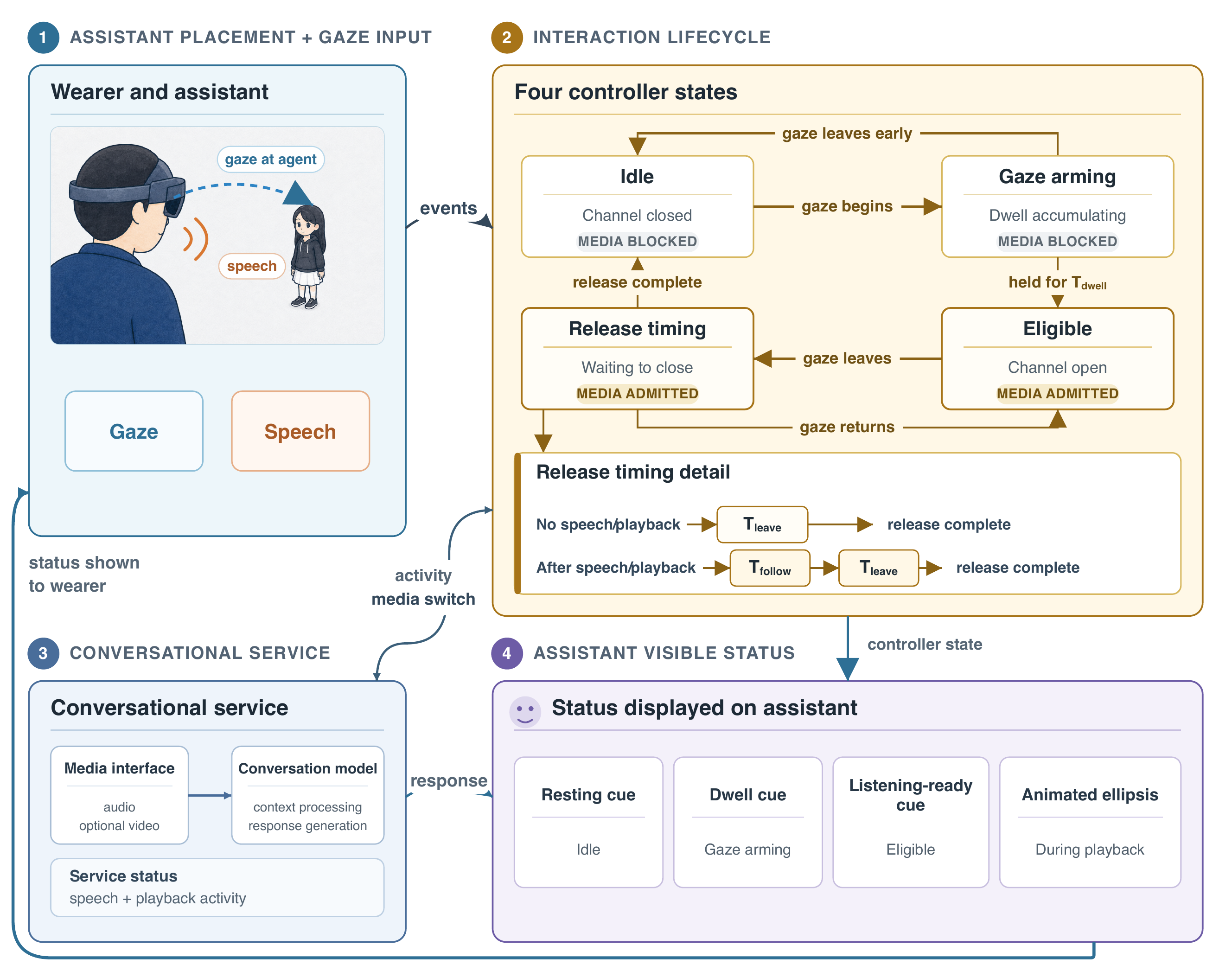}};
      \begin{scope}[x={(architecture.south east)},y={(architecture.north west)}]
        \fill[white] (0.430,0.392) rectangle (0.948,0.523);
        \node[anchor=west,text={rgb,255:red,31;green,44;blue,54},
          font=\sffamily\bfseries\fontsize{8.2}{9.2}\selectfont]
          at (0.435,0.507) {Release policy};
        \node[anchor=west,text={rgb,255:red,52;green,72;blue,91},
          font=\sffamily\fontsize{6.3}{7.2}\selectfont]
          at (0.435,0.445) {Retention depends on prior speech and playback activity.};
        \node[anchor=west,text={rgb,255:red,52;green,72;blue,91},
          font=\sffamily\fontsize{6.3}{7.2}\selectfont]
          at (0.435,0.415) {The exact timer sequence appears in Figure~\ref{fig:release-timing}.};
        \fill[white] (0.686,0.094) rectangle (0.797,0.143);
        \node[align=center,text={rgb,255:red,82;green,96;blue,109},
          font=\sffamily\fontsize{6.2}{7}\selectfont]
          at (0.7415,0.119) {Eligible and\\Release timing};
      \end{scope}
    \end{tikzpicture}
    \caption{System architecture and information flow. Gaze controls
    acquisition, speech and playback support retention, and inactivity leads to
    release. A local controller opens or closes the interaction channel, while
    the embodied assistant displays activation, listening, and response cues.
    The interaction-lifecycle panel shows the controller's four states and the
    release policy; Figure~\ref{fig:release-timing} details the timer sequence.}
    \label{fig:architecture}
\end{figure*}

\subsection{Design Requirements}

The target scenario combines three demands. First, users need a deliberate,
spatial way to address the assistant without confusing intentional selection
with incidental looking. Second, they must remain free to inspect materials or
look toward a collaborator after selecting the assistant. Third, they need to
know whether the assistant is still available for speech. These demands lead
to three design requirements: a stable embodied gaze target with an adjustable
activation dwell; retention that does not require continuous gaze; and visible
feedback for activation, listening, response activity, and release. Together, these
requirements treat invocation as a lifecycle rather than a single trigger.

\subsection{Interaction Lifecycle}

Figure~\ref{fig:architecture} follows one interaction from gaze acquisition to
release. A local controller combines gaze, speech activity reported by the
conversational service, and response playback on the headset. It has four
states: \emph{Idle}, \emph{Gaze arming}, \emph{Eligible}, and \emph{Release
timing}. The interaction channel is closed in Idle and Gaze arming and open in
Eligible and Release timing. In this paper, \emph{acquisition} denotes opening
the channel through sustained gaze, \emph{retention} denotes keeping that
channel open as attention moves elsewhere, and \emph{release}
denotes returning it to Idle.

\paragraph{Acquisition.}
From Idle, gaze entry starts the Gaze arming state and its timer
$T_\mathrm{dwell}$. Reaching the threshold opens the channel and moves the
controller to Eligible; looking away before the threshold returns it to Idle.
Dwell filters shorter incidental looks associated with the Midas-touch problem,
but its duration determines the trade-off between responsiveness and accidental
activation
\cite{majaranta2009fast,isomoto2022interaction,fernandez2020gazewheels,
namnakani2023comparing}. The threshold is therefore adjustable rather than
fixed as an intrinsic property of the interaction technique.

\paragraph{Retention.}
Direct gaze keeps the controller in Eligible. When gaze leaves, the controller
enters Release timing but keeps the channel open. If neither speech nor response
playback has occurred since acquisition, gaze departure starts the leave delay
$T_\mathrm{leave}$. After speech or playback, activity itself postpones release;
once both have ceased, the controller waits $T_\mathrm{follow}$ and then
$T_\mathrm{leave}$. New speech activity or playback restarts this post-activity
wait. Returning gaze moves the controller back to Eligible and pauses the
pending release countdown; if gaze leaves again, the countdown resumes.

This policy lets gaze establish the addressee without requiring continuous eye
contact during speech. People routinely avert gaze while planning an utterance
or consulting shared material, and cognitive difficulty increases gaze
aversion
\cite{glenberg1998averting,bogels2015neural,andrist2014conversational}.
Tama likewise documented listening failures when users looked toward notes or
ceased maintaining gaze \cite{mcmillan2019Tama}. Speech and playback therefore
retain an already-open channel but cannot open one themselves, consistent
with work showing that gaze, speech, feedback, and timing jointly organize
turns and joint attention
\cite{skantze2021turntaking,skantze2014jointattention,andrist2017looking}.

\paragraph{Release.}
The channel closes only when the applicable release countdown completes with
gaze absent and no new activity. The controller then returns to Idle, preventing
subsequent speech to the collaborator from being treated as assistant-directed.
Acquisition and release consequently occur under different attention
conditions: the user necessarily sees the assistant during gaze acquisition
but may be looking elsewhere when release occurs.

\subsection{Embodied Target and Visible Status}

\begin{figure}[H]
    \centering
    \includegraphics[width=\columnwidth]{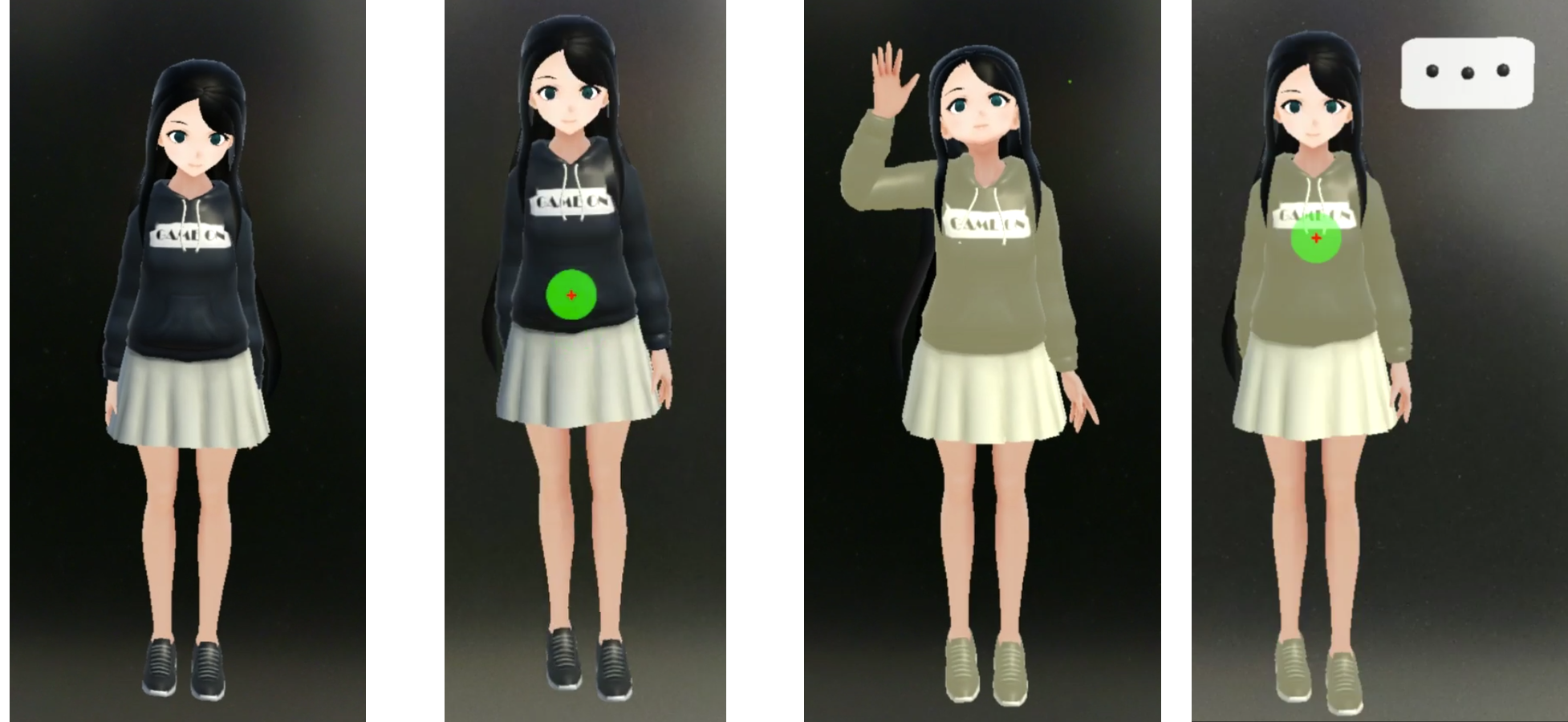}
    \caption{Embodied feedback during one activation cycle. From left to
    right: the resting cue, dwell cue, listening-ready cue, and animated
    ellipsis. The ellipsis appears during assistant-audio playback
    within the existing controller states.}
    \label{fig:embodied-feedback}
\end{figure}

The embodied assistant connects spatial address and system feedback. It gives
gaze a stable target and places activation progress and channel status on the
addressee itself
\cite{cassell2001embodied,kim2018body,wang2019exploring,
reinhardt2020embedding,urakami2023nonverbal}. The assistant is anchored at a
user-adjustable location beside the work area. Users can reposition it through
direct pinch or gaze-assisted pinch interaction, and it continuously rotates
around its vertical axis to face the wearer. Figure~\ref{fig:embodied-feedback}
shows the visible cues across one activation cycle.

\paragraph{Character form, scale, and orientation.}
The deployed assistant is a stylized, female-presenting humanoid approximately
20~cm tall. Its rig supplied legible idle, attentive, greeting, and response
poses; the miniature scale limited competition with the collaborator and
tabletop materials. Prior work shows that AR agent form and scale affect
perception and task fit
\cite{wang2019exploring,reinhardt2020embedding,hajahmadi2025voice,
schlager2024nonhumanoid}. Facing the wearer keeps its body cues and gaze target
visible; orientation also supports attention and coordination in embodied interaction
\cite{admoni2017social,urakami2023nonverbal,andrist2017looking}.

\paragraph{State feedback.}
Each controller state selects a visible cue.
Idle uses the resting cue. Gaze arming moves the character into an attentive
pose and displays dwell progress. Entering Eligible triggers a brief wave and a
listening-ready glow, which remains visible through Eligible and Release timing.
Response playback adds an animated ellipsis without introducing another
controller state. The assistant therefore presents activation progress,
channel-open status, and response activity at the same location used for
spatial address.

Colocating target and status makes state changes readily inspectable during
acquisition because the user is already looking at the assistant. It also
creates a visibility boundary: during retention and release, attention may have
returned to the materials or collaborator. Prior work examines peripheral,
audio, and nonverbal cues for maintaining awareness beyond the current visual
focus
\cite{hausen2014peripheral,hausen2014evaluating,klauck2017noticeable,
andrist2017looking}. Our evaluation therefore follows the interaction through
release rather than treating successful activation as the endpoint.

\section{Implementation}

This section describes how the interaction design was realized as a deployed
prototype. It locates the runtime components and communication paths, explains
how the acquisition--retention--release policy is enforced, and describes the
configuration interface. The implementation keeps time-critical
gaze decisions and visible state on the headset while using a remote service
for conversational processing.

\subsection{Runtime Deployment and Data Paths}

We implemented the system as a Unity~2022.3.62f1 application for Microsoft
HoloLens~2. Gaze sensing, interaction control, media gating, embodied feedback,
and response playback all run on the headset. The headset captures the wearer's
microphone audio and optional camera frames, while a remote conversational
service processes admitted speech and context and generates responses. A
persistent WebSocket session carries mono pulse-code-modulation (PCM) audio and
optional camera frames to the service, and returns service activity events and
response audio. The interaction controller decides locally whether to admit
media, before any user media enters the remote service.

A separate control path connects a companion application on the workstation to
the headset. It carries runtime configuration and explicit activation or
deactivation commands, while a telemetry path returns structured controller
records. User media travels only between the headset and conversational
service; it does not pass through the workstation. This separation supports
live configuration and logging without placing the workstation in the
user-media path.

\subsection{Local Interaction Control and Media Gating}

The headset queries the Mixed Reality Toolkit 3 (MRTK3) eye-gaze ray once per
Unity update and raycasts it against a capsule collider that follows the
assistant's visible body. At the deployed 0.15 scale, the collider measured
approximately 0.27~m high with a 0.045-m radius. The controller records gaze
entry when the ray begins intersecting the assistant and gaze departure when it
ceases; a ray miss is treated as departure.
The local interaction controller accumulates activation dwell and applies the
acquisition, retention, and release transitions in
Figure~\ref{fig:architecture}. It controls the current embodied cue and whether
the conversational media path is open.

The conversational service session remains connected while the controller
opens and closes a local media gate. A closed gate suppresses outgoing
microphone audio and optional camera frames, discards newly received response
audio, stops playback, and clears the playback buffer. Reopening the gate makes
the existing session available for a later engagement. While the channel is
open, recent speech activity reported by the service and local response
playback or buffering postpone release after gaze departure. Outgoing
microphone packets are withheld during response playback to reduce acoustic
feedback into the service; consequently, this prototype does not support user
barge-in during assistant playback. Together, these mechanisms implement the
controller states without coupling channel opening and closing to network
reconnection.

The service reports detected speech through asynchronous input-transcription
callbacks, and the controller treats speech activity as current for 1.5~s
after the most recent callback. Figure~\ref{fig:release-timing} contrasts the
two release paths. If neither a callback nor playback has occurred since
acquisition, gaze departure starts $T_\mathrm{leave}$ directly. After speech or
playback, release waits until both the speech-activity window and playback have
ended, then applies $T_\mathrm{follow}$ followed by $T_\mathrm{leave}$. With gaze
continuously absent and no new activity, the earliest return to Idle after a
final callback is therefore approximately
$1.5+T_\mathrm{follow}+T_\mathrm{leave}$ seconds; ongoing playback extends this
interval.

\begin{figure*}[!t]
    \centering
    \begin{tikzpicture}[
      state/.style={draw={rgb,255:red,70;green,91;blue,103},rounded corners=2pt,
        minimum height=8mm,align=center,font=\small,inner xsep=5pt},
      source/.style={state,fill=paperbluepale},
      wait/.style={state,fill=paperamber},
      closed/.style={state,fill=papergray},
      flow/.style={-{Latex[length=2mm]},line width=0.6pt,
        draw={rgb,255:red,70;green,91;blue,103}}
    ]
      \node[anchor=west,font=\small\bfseries] at (0,3.45)
        {No speech or playback since acquisition};
      \node[source,text width=3.0cm] (direct-start) at (1.8,2.75)
        {Gaze leaves\\\emph{Eligible}};
      \node[wait,text width=3.0cm] (direct-wait) at (7.0,2.75)
        {Wait $T_{\mathrm{leave}}$};
      \node[closed,text width=3.0cm] (direct-idle) at (12.2,2.75)
        {Idle\\channel closed};
      \draw[flow] (direct-start) -- (direct-wait);
      \draw[flow] (direct-wait) -- (direct-idle);

      \node[anchor=west,font=\small\bfseries] at (0,1.75)
        {After detected speech or response playback};
      \node[source,text width=2.7cm] (activity-start) at (1.6,1.05)
        {Gaze absent\\channel open};
      \node[source,text width=3.3cm] (activity-end) at (5.3,1.05)
        {Wait until speech window\\and playback end};
      \node[wait,text width=2.2cm] (follow) at (8.9,1.05)
        {$T_{\mathrm{follow}}$ wait};
      \node[wait,text width=2.0cm] (leave) at (11.8,1.05)
        {$T_{\mathrm{leave}}$ wait};
      \node[closed,text width=2.3cm] (activity-idle) at (14.8,1.05)
        {Idle\\channel closed};
      \draw[flow] (activity-start) -- (activity-end);
      \draw[flow] (activity-end) -- (follow);
      \draw[flow] (follow) -- (leave);
      \draw[flow] (leave) -- (activity-idle);
      \node[align=center,font=\scriptsize,text={rgb,255:red,52;green,72;blue,91}]
        at (5.3,0.33)
        {Speech activity remains current for 1.5~s after the final callback;\\
         playback can extend this period.};
      \node[align=center,font=\scriptsize\bfseries,
        text={rgb,255:red,128;green,63;blue,146}] at (8.2,-0.32)
        {Returning gaze restores \emph{Eligible}; new speech or playback restarts the post-activity wait.};
    \end{tikzpicture}
    \caption{Release timing while gaze remains absent. Before any speech or
    playback, gaze departure starts $T_\mathrm{leave}$ directly. After detected
    speech or response playback, the controller waits until both the 1.5-s
    speech-activity window and playback have ended, then applies
    $T_\mathrm{follow}$ and $T_\mathrm{leave}$. Returning gaze restores
    \emph{Eligible}; new speech or playback restarts the post-activity wait.}
    \label{fig:release-timing}
\end{figure*}

\subsection{Runtime Configuration and Service Interface}

Through the companion application, experimenters can update spatial offsets,
gaze-following thresholds, display settings, activation dwell, and the leave
wait. They can also start or stop a runtime stage, explicitly deactivate the
assistant, and enable or disable gaze processing. Parameter changes take effect
without rebuilding or restarting the headset application.

The headset communication layer presents the local controller with a stable
set of operations: maintain a session, admit or block media, expose recent
speech and playback activity, and deliver response audio. Service-specific
connection and message details remain within this layer.

\section{Experimental Setup and Calibration}

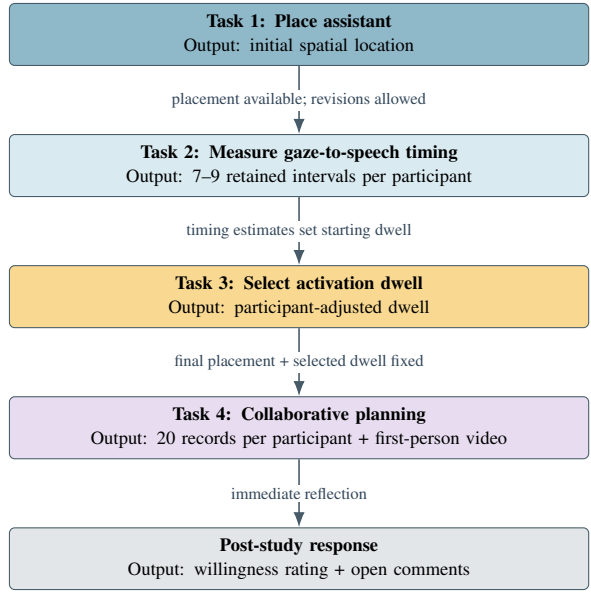
\begin{figure}[!t]
    \centering
    \begin{tikzpicture}[
      stage/.style={draw={rgb,255:red,70;green,91;blue,103},rounded corners=2pt,
        text width=0.88\columnwidth,align=center,inner ysep=4pt,font=\small},
      link/.style={-{Latex[length=2mm]},line width=0.6pt,
        draw={rgb,255:red,70;green,91;blue,103}},
      transfer/.style={fill=white,inner sep=1pt,font=\fontsize{6.2}{7}\selectfont,
        text={rgb,255:red,52;green,72;blue,91}}
    ]
      \node[stage,fill=paperblue] (task1)
        {\textbf{Task 1: Place assistant}\\Output: initial spatial location};
      \node[stage,fill=paperbluepale,below=9mm of task1] (task2)
        {\textbf{Task 2: Measure gaze-to-speech timing}\\
         Output: 7--9 retained intervals per participant};
      \node[stage,fill=paperamber,below=9mm of task2] (task3)
        {\textbf{Task 3: Select activation dwell}\\
         Output: participant-adjusted dwell};
      \node[stage,fill=paperpurple,below=9mm of task3] (task4)
        {\textbf{Task 4: Collaborative planning}\\
         Output: 20 records per participant + first-person video};
      \node[stage,fill=papergray,below=9mm of task4] (post)
        {\textbf{Post-study response}\\
         Output: willingness rating + open comments};
      \draw[link] (task1) -- node[transfer] {placement available; revisions allowed} (task2);
      \draw[link] (task2) -- node[transfer] {timing estimates set starting dwell} (task3);
      \draw[link] (task3) -- node[transfer] {final placement + selected dwell fixed} (task4);
      \draw[link] (task4) -- node[transfer] {immediate reflection} (post);
    \end{tikzpicture}
    \caption{Study workflow and outputs at each stage. Tasks~1--3 established
    participant-specific placement and dwell. Task~4 used that configuration
    during collaborative planning and produced 20 chronological records per
    participant plus first-person video for review, followed by an immediate
    post-study response.}
    \label{fig:study-workflow}
\end{figure}

Before the collaborative evaluation, three tasks set the assistant location
and activation dwell used by each participant. Short dwell speeds activation
but admits more incidental looks; long dwell filters those looks but delays access
\cite{majaranta2009fast,isomoto2022interaction,
fernandez2020gazewheels,mohan2018dualgaze,pfeuffer2020empirical}.
Figure~\ref{fig:study-workflow} summarizes the complete procedure and the
configuration or evidence carried from each stage to the next.

\subsection{Participants, Ethics, and Procedure}

We recruited a 7-person engineering pilot to refine an earlier
implementation and a formal cohort of 25 participants for the reported study.
All pilot data are excluded from the analyses and figures. The study protocol
was reviewed and approved through the authors' institutional ethics process,
and all participants provided informed consent covering participation,
first-person recording, voice processing, and de-identified research use.
Formal cohort study logs use de-identified identifiers (IDs). The retained recruitment record
was audited against the local participant register to link those IDs before
aggregation. Formal participants were 23--36 years old (mean 27.0, standard
deviation (SD) 3.6);
14 identified as men, 10 as women, and 1 preferred not to report gender.
Education levels were bachelor's (3), master's (14), and doctoral (8). Vision
was normal without correction for 6 participants, corrected with glasses for
14, and corrected with contact lenses for 5. Prior experience included mobile
AR for 20 participants, a VR headset for 22, and a head-worn AR or
mixed-reality (MR) device for 12. Preferred study languages were Chinese (17),
Japanese (7), and English (1); task materials were localized.

Every formal participant completed the sequence in
Figure~\ref{fig:study-workflow} in the same order.

\subsection{Apparatus and Experimental Configuration}

Participants completed the HoloLens~2 eye calibration before the study. The
formal tasks used the prototype described in Section~4 with ByteDance
Doubao Realtime Dialogue over WebSocket. The headset captured 16-kHz, 16-bit
mono PCM audio in 1024-sample chunks; camera transmission to the service was
disabled. The headset used wearer-focused microphone processing, but the study
did not independently measure leakage of collaborator speech into capture.
Character style, scale, and orientation policy remained constant across the
formal tasks.

The companion workstation applied the parameters selected during Tasks~1--3
and received placement, gaze-to-speech timing, dwell, and controller records.
For study instrumentation, the HoloLens displayed a persistent gaze cursor---a
red heart with a small green crosshair-like ring. It followed the gaze locus and
remained distinct from the assistant's state cues. First-person recording
preserved this cursor, the assistant cues, and the surrounding activity. Task~4
used each participant's final dwell value.
The follow-up and leave waits were fixed at 1.0~s for every session; the
controller's 1.5-s speech-activity window remained unchanged. Thus all
participants experienced the same release timing after calibration.

\subsection{Task 1: Assistant Placement}

\begin{figure}[!t]
    \centering
    \includegraphics[width=0.98\columnwidth]{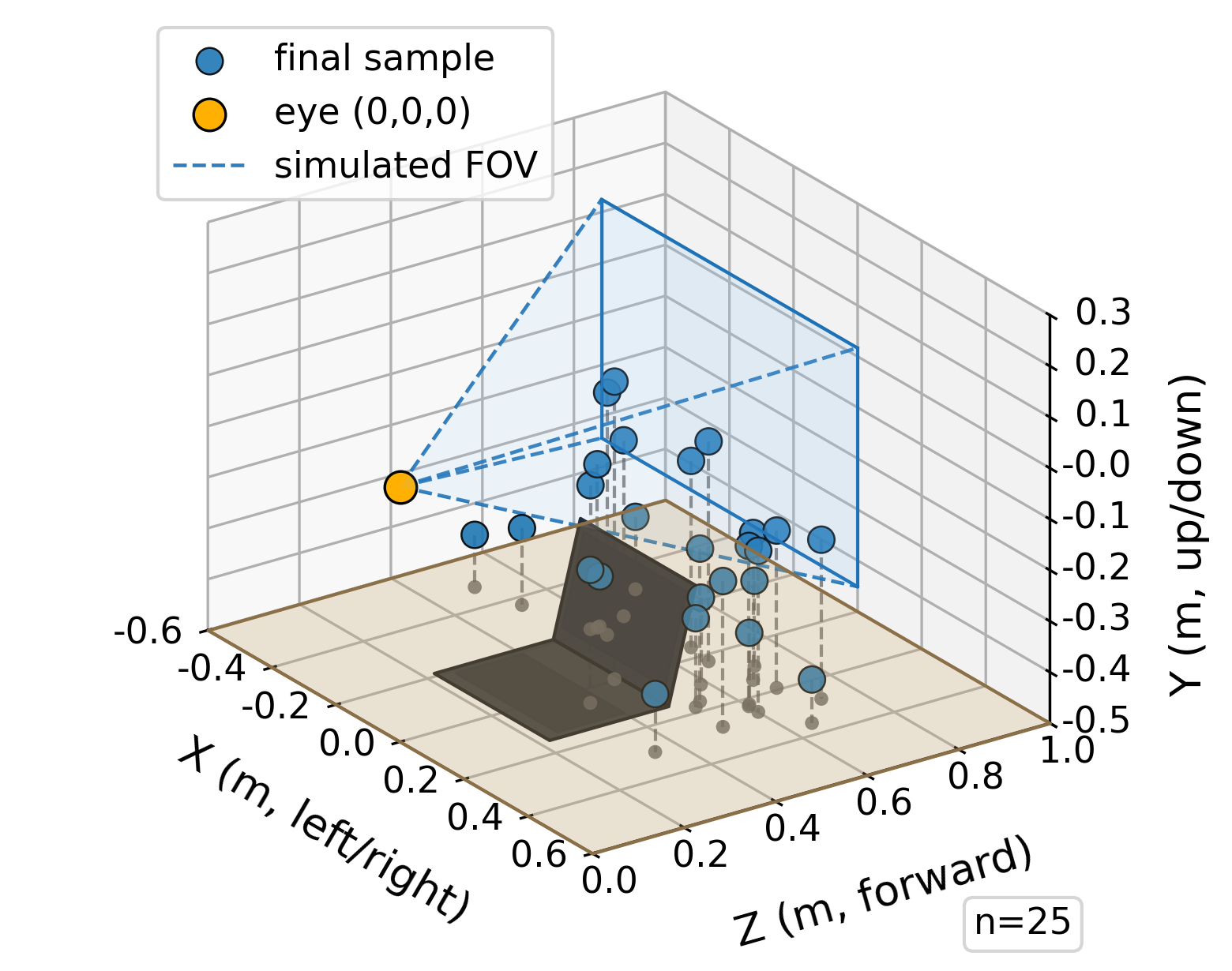}
    \\[-1mm]
    \includegraphics[width=0.98\columnwidth]{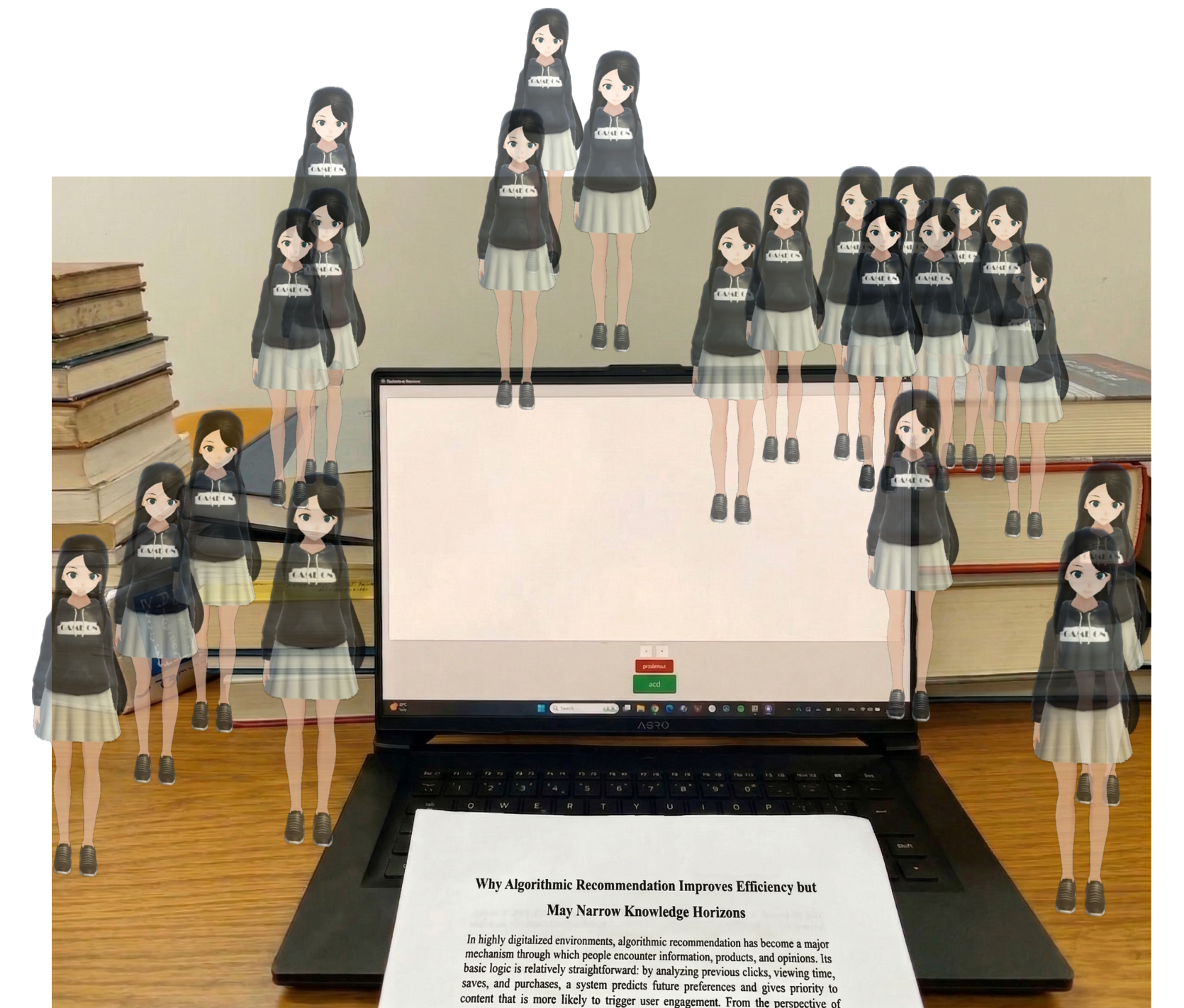}
    \caption{Final Task~1 placements for the formal cohort ($n=25$). Top: the
    three-dimensional distribution relative to the headset, simulated field of
    view, and tabletop reference. Bottom: the placements overlaid on the desk
    scene. Most selections were near the task materials and below eye level,
    with greater lateral than distance variation.}
    \label{fig:task1-position}
\end{figure}

Task~1 let participants place the assistant where it remained easy to address
without obstructing the collaborative work area; Figure~\ref{fig:task1-position}
shows the final selections. Audio interaction was disabled, while dwell and
listening-ready cues remained available for familiarization. A fixed 0.8~s
dwell provided the same placement condition for all participants; this value
lies within durations examined in prior gaze-input studies
\cite{majaranta2009fast,isomoto2022interaction,
fernandez2020gazewheels,namnakani2023comparing}. Participants worked with a
laptop and paper materials, repositioned the assistant, and confirmed a
location. The system recorded the center of its rendered mesh bounds relative
to the headset. Participants could revise the placement during Task~2.

For analysis, we retained the last confirmed placement for each participant.
Final placements were relatively consistent in distance and varied in
horizontal direction: distance had mean/median 0.685/0.68~m (interquartile range (IQR)
0.64--0.74~m), azimuth had mean/median $2.9^\circ$/$8.3^\circ$ (IQR
$-19.1^\circ$ to $20.2^\circ$), and elevation had mean/median
$-20.3^\circ$/$-19.9^\circ$ (IQR $-29.1^\circ$ to $-15.0^\circ$).
Most participants placed the assistant below eye level and near the work
materials, with greater horizontal than distance variation. This pattern
supports an initial location near the materials while preserving horizontal
adjustment for the user.

\subsection{Task 2: Gaze-to-Speech Timing}

Task~2 measured the interval from gaze onset to speech onset. The assistant was
force-activated so that visible state changes would not cue when participants
began speaking. Participants received reading materials and were asked to
interact with the assistant. Their only instruction was to look at the
assistant before speaking; they chose when to speak and what to ask. The
controller timestamped gaze onset and,
after the conversational service reported speech, searched backward through a
20-Hz history of local microphone root-mean-square (RMS) values to estimate the
earlier speech onset.

The detector retained 10~s of pre-gain RMS values, required a local peak of at
least 0.02, and treated values below 0.01 as silence. Its 3.0~s initial lookback
adapted after 3 service notifications to the recent 80th-percentile delay
plus 0.7~s, bounded to 1.5--8.0~s and the available history. We selected these
thresholds during development by comparing the stored waveforms with the
corresponding audible utterances. The resulting onset is therefore a locally
calibrated estimate rather than the later service notification.

\begin{figure}[!t]
    \centering
    \includegraphics[width=\columnwidth]{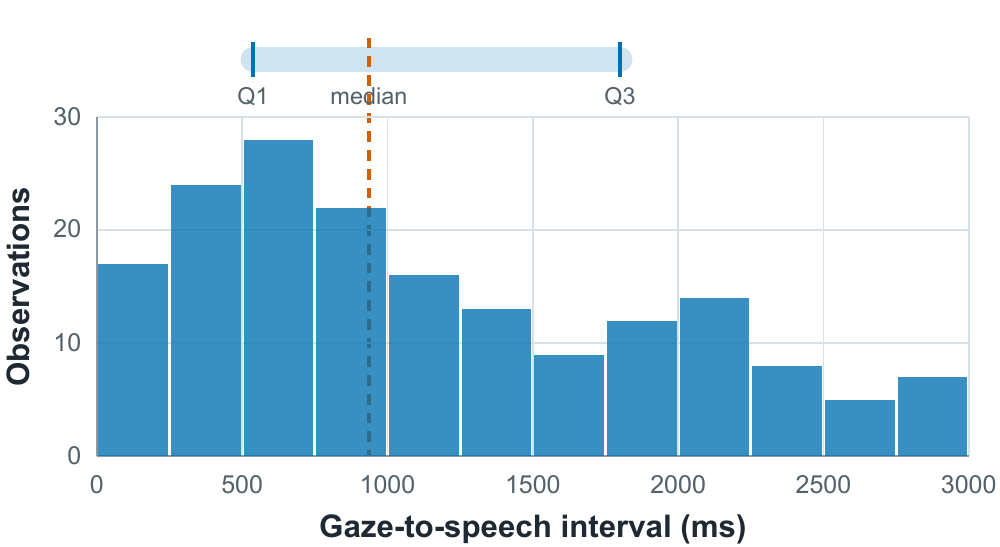}
    \caption{Task~2 gaze-to-speech timing for the formal cohort (175 retained
    observations; first 7 per participant). The band marks the IQR, and the
    dashed line marks the median. The figure shows the equal-participant-weight
    subset; Task~3 used all retained measurements available for each participant.}
    \label{fig:task2-timing}
\end{figure}

The controller stored a measurement only when the estimated gaze-to-speech
interval was within 100--3000~ms. The formal logs contained 178 retained
measurements: 23 participants contributed 7, 1 contributed 8, and 1 contributed
9. To give each participant equal weight, the distribution analysis used the
first 7 retained measurements per participant, yielding 175 observations. The
corresponding service-notification lag had median 770~ms (mean 827.1~ms; range
0--4666~ms). Figure~\ref{fig:task2-timing} shows that gaze-to-speech timing
varied substantially: mean 1155.8~ms
(SD 774.6~ms), median 936~ms, IQR 536.5--1800~ms, and range 100--2934~ms.
Because rejected episodes were discarded before logging, their number is
unavailable; these statistics describe the retained measurements rather than
every gaze episode observed during Task~2.

During the sessions, experimenters qualitatively noted shorter intervals for
greetings and routine questions and longer pauses when participants appeared to
formulate more complex requests. This pattern is compatible with evidence that
response planning starts once relevant content becomes available and that
speech formulation requires several hundred milliseconds
\cite{bogels2015neural}; cognitive difficulty also systematically changes gaze
behavior \cite{glenberg1998averting}. The observed variation made the measured
interval useful as a starting value, followed by explicit adjustment in
Task~3.

\subsection{Task 3: Dwell Selection}

\begin{figure}[!t]
    \centering
    \includegraphics[width=\columnwidth]{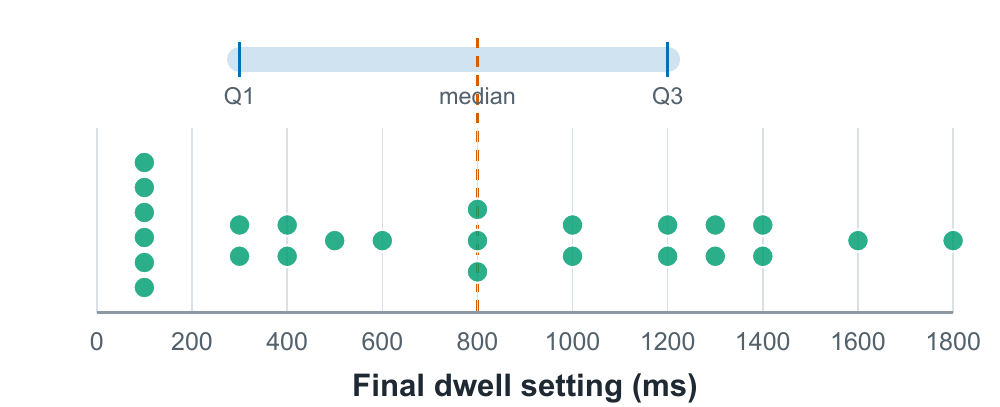}
    \caption{Final Task~3 dwell settings for the formal cohort ($n=25$). Each
    dot is one participant's final explicit choice; the band marks the IQR and
    the dashed line marks the median.}
    \label{fig:task3-dwell}
\end{figure}

At the start of Task~3, the runtime used all retained Task~2 measurements then
available for that participant (7--9 measurements). To reduce the influence of
an unusually short or long interval on the starting estimate, it removed the
fastest and slowest values, averaged the remainder, rounded to the nearest
100~ms, and clamped the result to 100--2000~ms. The resulting starting dwell had
mean 1088~ms, median 1100~ms, and range 500--2000~ms. Participants then tried
the gaze trigger and adjusted this starting value in 100-ms steps using the
embodied target and cues, a process
consistent with prior gaze-dwell calibration
\cite{majaranta2009fast,isomoto2022interaction,fernandez2020gazewheels,
pfeuffer2020empirical}. The selected value was fixed for Task~4. Final dwell
ranged 100--1800~ms (mean 748.0, SD 543.2, median 800, IQR 300--1200~ms;
Figure~\ref{fig:task3-dwell}). Of the 25 participants, 19 reduced the starting
value, 5 increased it, and 1 retained it; the median change was $-400$~ms. The
calculated value served only as an initial setting for adjustment and was not
treated as an optimized dwell threshold.

\section{Collaborative Interaction Test and Event Analysis}

\begin{figure*}[!t]
    \centering
    \begin{tikzpicture}
      \node[anchor=south west,inner sep=0] (outcomes) at (0,0)
        {\includegraphics[width=\textwidth]{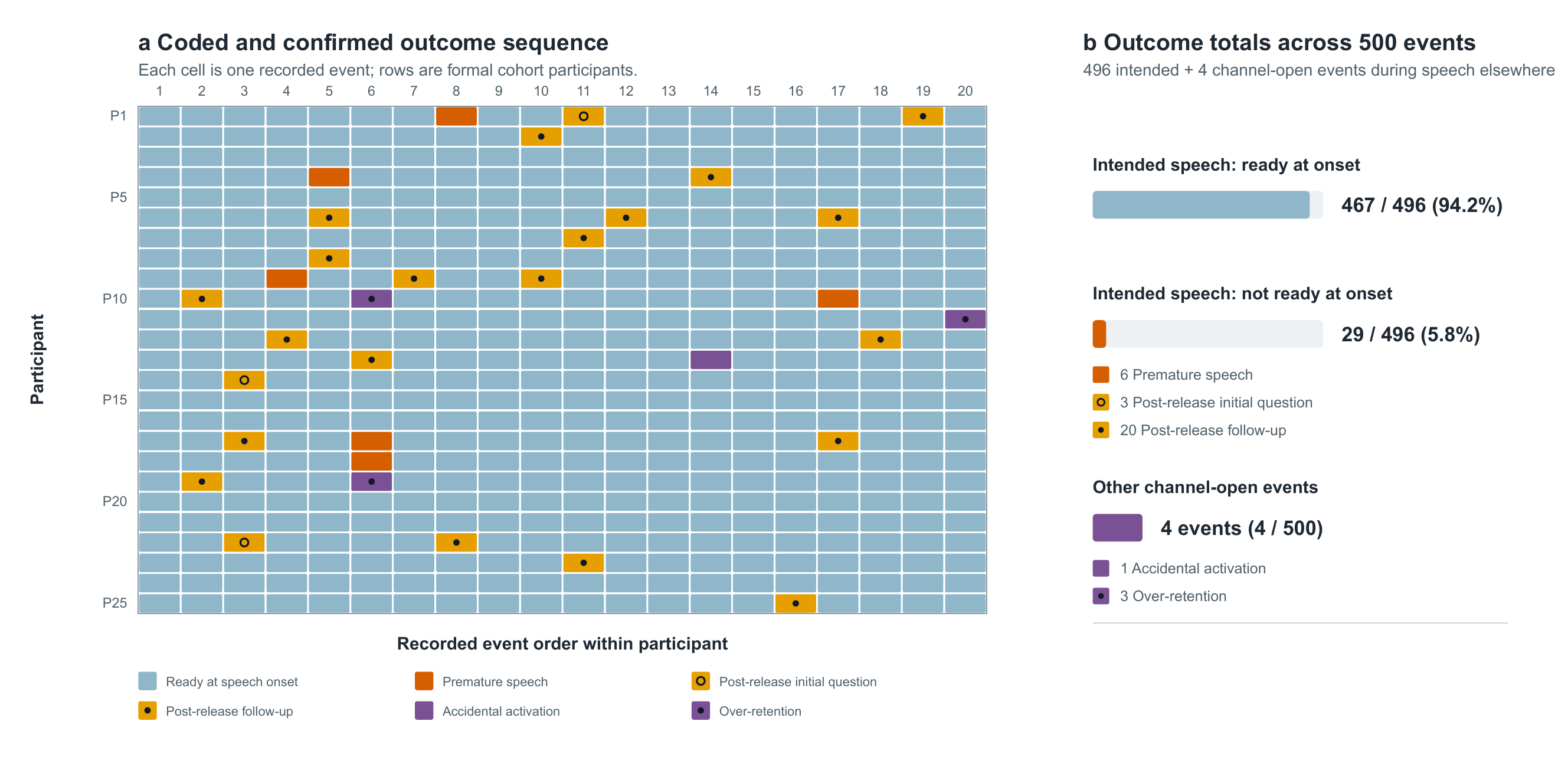}};
      \begin{scope}[x={(outcomes.south east)},y={(outcomes.north west)}]
        \fill[white] (0.689,0.891) rectangle (1.000,0.929);
        \node[anchor=west,text={rgb,255:red,82;green,97;blue,107},
          font=\sffamily\fontsize{5.5}{6.5}\selectfont]
          at (0.691,0.901) {496 intended + 4 other channel-open events};
      \end{scope}
    \end{tikzpicture}
    \caption{Task~4 outcomes across 25 participants (20 sequentially recorded
    events each; 500 total). Outcome terms match Table~\ref{tab:task4-terms};
    colors and symbols distinguish Premature speech (dark orange), Post-release
    initial question (ringed amber), Post-release follow-up (dotted amber),
    Accidental activation (solid purple), and Over-retention (dotted purple).}
    \label{fig:task4-outcomes}
\end{figure*}

\subsection{Collaborative Planning Task}

Task~4 evaluated the configured assistant in a collaborative tabletop
travel-planning activity. Each participant worked with a researcher acting as
the collaborator. The researcher gave no instructions or hints about assistant
use and did not attempt to produce particular events; they conversed naturally
as the participant's planning partner. Together they consulted printed
materials, discussed constraints, queried the assistant, and recorded their decisions. Gaze and speech moved
between several plausible targets, as in multiparty turn coordination
\cite{auer2018gaze,bohus-horvitz-2011-multiparty,andrist2017looking,
lee2024may}. The assistant was visible to the participant but not to the
collaborator, who followed the same planning prompt and did not operate the
interface.

Each participant used the dwell selected in Task~3. Reading, writing, and
conversation repeatedly shifted attention and speech among the assistant,
materials, and collaborator, testing the configured interaction during natural
task activity. A staff member recorded
events in chronological order, and the task ended when the live counter reached
20 records. All formal logs contain indices 1--20 for each participant. The
resulting fixed sample contains 496 intended assistant-directed utterance
attempts and four other channel-open events.
Continuous first-person recording preserved the surrounding context. After the
20th record, participants answered open-ended questions about the interaction
and possible improvements.

\subsection{Event Recording and Critical-Incident Coding}

We combined fixed event recording with critical-incident review. Critical
Incident Technique focuses analysis on concrete episodes that reveal
requirements for effective activity \cite{flanagan1954critical}. HCI studies of
communication systems similarly combine observation with post-session accounts
to derive design requirements \cite{kies2000augmenting}, while first-person
video preserves the temporal relation among gaze, speech, body movement, and
system feedback \cite{webb2018video}.

During Task~4, a staff member outside the participant's view entered a live
marker for each recorded event while watching the first-person perspective.
This staff member was not the collaborator. The runtime interface provided 3
provisional categories: successful activation, failed activation, and apparent
unintended activation. These markers established the chronological event sample
but did not determine the final classifications.

Two researchers subsequently reviewed every complete first-person recording
and reached consensus on the final code. Table~\ref{tab:coding-map} shows how
the live markers were resolved through this review. The visible gaze cursor and
assistant cues allowed the researchers to reconstruct gaze entry and departure,
system state, speech onset, and intended addressee; participants clarified
ambiguous addressee intent when needed. Every chronological record ultimately
received one of the fixed terms in Table~\ref{tab:task4-terms}.

\begin{table}[H]
\caption{Mapping from provisional live markers to final video-review codes.}
\label{tab:coding-map}
\centering
\scriptsize
\setlength{\tabcolsep}{3pt}
\renewcommand{\arraystretch}{1.08}
\begin{tabular}{@{}p{0.24\columnwidth} c p{0.58\columnwidth}@{}}
\hline
\textbf{Live marker} & \textbf{$n$} & \textbf{Final classification} \\
\hline
\shortstack[l]{Successful\\activation} & 467 & Ready at speech onset (467) \\
\shortstack[l]{Failed\\activation} & 29 & Premature speech (6); Post-release initial question (3); Post-release follow-up (20) \\
\shortstack[l]{Apparent\\unintended\\activation} & 4 & Accidental activation (1); Over-retention (3) \\
\hline
\end{tabular}
\end{table}

For every assistant-directed utterance attempt, reviewers coded whether the
interaction channel was open or closed at speech onset. To examine retention,
they also identified attempts in which gaze left the assistant after acquisition
and before speech. Synchronized runtime diagnostics confirmed that visible
state changes matched channel opening and closing. Automatic speech
recognition (ASR) content, response
quality, and planning-task success were outside this analysis. Task~4 was
designed to locate outcomes within one integrated acquisition--retention--release
policy, not to compare gaze with wake words or isolate the causal contribution
of each component. This use of a deployed artifact to produce design knowledge
follows a research-through-design approach \cite{zimmerman2007research}.

\subsection{Outcomes Across Acquisition, Retention, and Release}

\begin{table}[!t]
\caption{Final classification of the 500 Task~4 outcomes.}
\label{tab:task4-terms}
\centering
\scriptsize
\setlength{\tabcolsep}{3pt}
\renewcommand{\arraystretch}{1.08}
\begin{tabular}{@{}l c p{0.55\columnwidth}@{}}
\hline
\textbf{Term} & \textbf{$n$} & \textbf{When it occurred} \\
\hline
Ready at speech onset & 467 & Intended speech began while the assistant was listening. \\
Premature speech & 6 & The participant looked at the assistant but spoke before dwell completed and before the ready cue appeared. \\
\shortstack[l]{Post-release initial\\question} & 3 & After activation, the participant looked back to the materials to prepare the first question; the assistant returned to Idle before they spoke. \\
Post-release follow-up & 20 & After an assistant-directed turn, the participant prepared a follow-up; the assistant returned to Idle before they spoke again. \\
Accidental activation & 1 & The assistant activated although the participant did not intend to address it. \\
Over-retention & 3 & After an assistant-directed turn, the participant turned and spoke to the collaborator before the channel closed. \\
\hline
\end{tabular}
\end{table}

Figure~\ref{fig:task4-outcomes} and Table~\ref{tab:task4-terms} summarize the
fixed sample of 500 recorded outcomes. It comprised 496 assistant-directed
utterance attempts and 4 other channel-open events. Of the 496 intended
attempts, 467 (94.2\%) were \emph{Ready at speech onset}; the other 29 began
while the channel was closed. Gaze acquisition completed before speech in 490
attempts (98.8\%): the 467 ready attempts and 23 post-release events. The remaining 6
were \emph{Premature speech} events in which speech began before dwell
completed. Thus, 98.8\% describes acquisition completed before speech, not ASR
accuracy, response quality, or task success.

The 6 premature-speech events came from 6 participants. The 23 post-release
events comprised 3 \emph{Post-release initial question} and 20
\emph{Post-release follow-up} events and spanned 16 participants. In each, the
assistant had previously activated and returned to Idle while the participant
looked elsewhere; the participant then addressed the assistant without
reactivating it. Altogether,
17 participants contributed at least 1 intended utterance that began while the
channel was closed. Four participants contributed 19 intended attempts and 1
other channel-open event; the other 21 contributed 20 intended attempts. Across
participants, the median proportion of intended attempts that were ready at
speech onset was 95\% (range 85--100\%).

Consensus video review identified 102 intended attempts in which gaze left
after acquisition but before speech. The configured retention policy kept the
channel open for 79 (77.5\%); the other 23 (22.5\%) were the post-release events
defined above. This percentage is conditional on those 102 reviewed
gaze-departure sequences, rather than a global retention-success rate.

The remaining 4 outcomes comprised 1 \emph{Accidental activation} and 3
\emph{Over-retention} events. These counts distinguish an accidental opening
from a channel that remained open as speech shifted to the collaborator.
Because the fixed 20-event sample does not enumerate all incidental looks or
all collaborator-directed utterance opportunities, these counts are not
false-activation or over-retention probabilities.

\subsection{Post-Study Interview and Willingness}

Immediately after Task~4, participants rated their willingness to use
gaze-based invocation, setting aside HoloLens weight and response quality. Of
the 25 participants, 10 selected strongly willing, 11 somewhat
willing, and 4 neutral; none selected an unwilling option. Open-ended material
was used as descriptive context rather than as a systematically coded
qualitative dataset, so we make no prevalence claims from it. Comments described
gaze-based invocation as convenient and activation status as clear, including one description
of the activated and non-activated states as ``very conspicuous'' (authors'
translation). Other comments mentioned needing to look at task materials and
learning a gaze-then-speak habit. One participant reported that input stopped
before an utterance ended; because this statement was not aligned to a specific
runtime record, we treat it as an illustrative perception rather than a
separately counted event.

\FloatBarrier

\section{Discussion}

\subsection{Acquisition through an Embodied Assistant with Visible Status}

Among the 496 assistant-directed utterance attempts, gaze acquisition completed
before speech in 490 (98.8\%); the other 6 (1.2\%) were \emph{Premature speech}
events. This is a within-system measure of whether dwell completed before
speech. It is not a measure of response correctness or task success, and the
study did not include a wake-word condition from which to infer relative
modality performance.

Participants looked toward the intended addressee, saw acquisition progress and
readiness on the same body, and then spoke. Debrief comments included descriptions
of activation status as clear and conspicuous. Together, deliberate gaze and an explicit visual state
change demonstrate the feasibility of using gaze as a spatial activation
signal while visible status lets users verify that the assistant is ready.

Embodiment served a functional role: the anchored character served as addressee,
interaction target, and status display, turning a hidden detector decision into
a visible interaction. Participant-specific dwell changes the required gaze
duration while preserving the same action and feedback across users.

\subsection{Retention, Release, and Later Intended Speech}

Within the 102 reviewed cases in which gaze left after acquisition but before
speech, the configured retention policy kept the channel open for 79 (77.5\%), letting
participants inspect materials or formulate a question without continuous
fixation. The other 23 were post-release events. In 3 \emph{Post-release
initial question} events, participants activated the assistant and then looked
at the materials to prepare their first question; the assistant returned to
Idle before they spoke. In 20 \emph{Post-release follow-up} events, participants
had completed an assistant-directed turn and were preparing the next question
when the assistant returned to Idle. These events spanned 16 participants.
Because all 102 reviewed events share the same order
of acquisition, gaze departure, and speech, they show how the configured timing
handled continued interaction and when it ended too soon.

The opening and closing states differ in what the user can see. During
acquisition, the wearer looks at the embodied character, where its attentive
pose, wave, glow, and the gaze cursor show the interaction state. After
attention returns to materials or a collaborator, the cursor follows the new
gaze location while the assistant's status cue remains on the assistant. The cue therefore
becomes spatially separated from the wearer's current focus.
Retention is consequently valuable because it lets users formulate language
while leaving gaze free to move. After release, users may attempt the next turn
while believing the assistant remains active because its status remains at the
prior gaze location.

Video confirmed the complete activation--release sequence for every
post-release event. The resulting design question is how to preserve
natural attention shifts while communicating that the assistant has stopped
listening when the user is looking elsewhere.

\subsection{Design Lessons for Gaze-Activated AR Assistants}

The first lesson is that embodiment serves two functions. In optical see-through AR, one
character anchored in the physical scene can be both a spatial addressee target and the place
where status is displayed, consistent with work on embodied form and nonverbal coordination
\cite{cassell2001embodied,kim2018body,wang2019exploring,
reinhardt2020embedding,urakami2023nonverbal,andrist2017looking}. The observed
acquisition pattern supports colocating address and status at the attended
target. Once attention moves, however, release feedback should reach the user
through a peripheral visual, audio, or haptic channel
\cite{hausen2014peripheral,klauck2017noticeable,lee2024may}.

The second lesson is to separate opening the channel from keeping it open. Gaze
establishes the addressee, while speech and playback keep the channel open as visual
attention returns to the task \cite{skantze2021turntaking,
skantze2014jointattention,andrist2017looking}. This retention supports natural
planning behavior without requiring every utterance to begin under direct
gaze, as demonstrated by the 79 events in which gaze left before speech while
the channel remained open. Release
then deserves an explicit cue available beyond the assistant's body, such as a short spatial sound, peripheral marker, or wearable haptic.

The third lesson is to expose and evaluate each controller transition.
Participant-selected dwell accommodates speed--error preferences, while fixed
release waits make the closing threshold explicit and inspectable
\cite{majaranta2009fast,fernandez2020gazewheels}. Future systems can test
adaptive retention using speaker-aware audio processing together with gaze
location, head orientation, and turn cues. Evaluations can replay critical
incidents and count all relevant
opportunities, preserving the transition that preceded
each outcome rather than reducing the interaction to one activation total.

\subsection{Scope and Limitations}

Task~4 recorded the first 20 chronologically marked events for each participant,
comprising assistant-directed utterance attempts and other channel-open events.
This fixed sampling gives each participant equal event weight, but it does not
enumerate every incidental gaze opportunity or every period of non-addressed
speech. The accidental-activation and over-retention findings are therefore
counts within these 500 records rather than time-normalized error rates.

The study evaluated one integrated configuration without a wake-word or
component-ablation condition. It can locate outcomes within acquisition,
retention, and release, but cannot establish superiority over wake words or
isolate the causal contributions of embodiment, visible status, personalized
dwell, and retention timing. The follow-up and leave waits were fixed across
participants, making release timing comparable but leaving parameter
optimization for future work. The prototype also withheld microphone packets
during assistant playback, so the findings do not cover barge-in interaction.

Task~2 retained only gaze-to-speech estimates between 100 and 3000~ms; rejected
episodes were not logged, so their number cannot be reconstructed. The
calculated dwell served only as a starting point for explicit participant
adjustment, not as a validated optimum. In addition, the visible gaze cursor
supported review but may itself have influenced gaze behavior and awareness of
system state.

The study recorded 20 events per participant in a single session. This
short exposure cannot show learning effects or how gaze and turn-taking habits
change through repeated use. One participant expected interaction to
become easier after learning the gaze-then-speak sequence. Longer deployments
can examine whether users adapt their timing or become more aware of release.

All participants were pursuing or had completed a bachelor's degree, and most
had prior extended reality (XR) experience, including mobile AR (20), VR headsets (22), or
head-worn AR/MR (12). Users with less formal education or no prior XR exposure
may develop different interaction strategies.

Finally, the gaze cursor, assistant cues, and synchronized diagnostics supported
reconstruction of the recorded event histories. The two researchers reached
consensus rather than producing independent reliability codes, so no inter-rater
reliability statistic is claimed. Participant clarification supported ambiguous
addressee judgments, while open-ended debriefs were used only as illustrative
context. The collaborative tabletop planning task, HoloLens~2 implementation,
and one conversational service define the current boundary of generalization.

\section{Conclusion}

We examined gaze-based interaction with an embodied AR assistant whose status
was visible to the user. Users opened the interaction by looking at the assistant and seeing
its ready state on the same body. Speech and playback then kept the channel open
as attention returned to the task. Participant-selected placement and dwell,
together with the collaborative test, showed that this design provided a
consistent entry into assistant interaction within the evaluated configuration.

Following the interaction beyond activation revealed an equally important
challenge at release. While planning what to say, users sometimes looked back to
task materials and could no longer see the assistant's status. Most intended
utterances with a closed channel occurred after an earlier activation and
release, rather than during the initial gaze trigger. The resulting design
lesson spans the full interaction lifecycle: embodied visible status clarifies
entry, retention supports natural shifts of visual attention, and release needs
cues that remain perceptible after attention moves elsewhere. Our implementation
and findings provide concrete guidance for future gaze-activated assistants.

\bibliographystyle{template/abbrv-doi}
\bibliography{references}

\end{document}